\documentclass[prl,showpacs,preprintnumbers]{revtex4}
\usepackage{amsfonts}
\usepackage{mdframed}
\usepackage{amssymb}
\usepackage{footnote}
\usepackage{amsmath}
\usepackage{graphicx}
\usepackage{float}
\usepackage[font={footnotesize,it}]{caption}
\usepackage[utf8]{inputenc}
\usepackage{natbib}
\usepackage{tikz}
\usepackage{tikz-3dplot}
\usepackage[colorlinks=true,
            linkcolor=purple,
            urlcolor=purple,
            citecolor=blue]{hyperref}

\begin{document}

\title{$n+1$-dimensional Bertotti-Robinson solutions in gravity coupled with
nonlinear electrodynamics}
\author{S. Habib Mazharimousavi}
\email{habib.mazhari@emu.edu.tr}
\author{M. Halilsoy}
\email{mustafa.halilsoy@emu.edu.tr}
\affiliation{Department of Physics, Faculty of Arts and Sciences, Eastern Mediterranean
University, Famagusta, North Cyprus via Mersin 10, Turkey}
\date{\today }

\begin{abstract}
We consider the pure electric type power-law Lagrangian in nonlinear
electrodynamics theory of dimensions $n+1$. Conformally flat form of
Bertotti-Robinson type spacetimes are obtained in all these dimensions with
the specific power $s=-\frac{1}{n-4},$which naturally excludes $n=4.$ This
particular dimension is considered separately which is not of power-law form.
\end{abstract}

\maketitle

The unique, source free, conformally flat (CF) non-null solution of the
Einstein-Maxwell (EM) theory in $3+1$-dimensions is known to be the
Bertotti-Robinson (BR) solution \citep{1,2}. Being a regular solution with a
uniform electromagnetic (em) field made the BR solution rather useful in
general relativity and particularly in string theory. In $n=3$ both the em
field and its dual are $2$-forms and this symmetry plays role in obtaining
the BR solution. An analogue solution in $n+1$-dimensions ($n\geq 2$),
however, has no such symmetry while going from the topology $AdS_{2}\times
S^{2}$ to $AdS_{2}\times S^{n-1}.$ We resort therefore, in this study, to
the nonlinear electrodynamics (NED) in search for such CF solutions in $n+1$%
-dimensions. CF spacetimes have properties in common with flat spacetimes
and in this sense they used to attract much interest. For example, besides
absence of conformal curvature the null geodesics in such spacetimes remain
undeflected as in the flat spacetime. Although NED deflects light this
doesn't occur when the underlying spacetime is CF. Our NED Lagrangian is of
the form $\mathcal{L}\sim \mathcal{F}^{s}$, where $\mathcal{F}=F_{\mu \nu
}F^{\mu \nu }$ is the Maxwell invariant and $s$ is a constant parameter as
power \citep{PL}. Such power-law Lagrangians have been considered in the
Literature \citep{3,4,5} to explain certain nonlinear features of
electromagnetism when coupled with gravitation. These can be summarized as
confinement of particle geodesics / charges, vortex solutions in string
theory, quantum wave probes against singularity, superconductivity,
condensation and other features that we hardly encounter in linear Maxwell
electrodynamics. We also refer to \cite{PLHD} (and the references therein)
for works where this type of Lagrangian has been analyzed, in different
aspects of general relativity. It is well known that the origin of NED in
field theory took start with the work of Born and Infeld \citep{6}, which
aimed to remove the singularities due to point charges. Let us add that the
model of Born and Infeld had the limit of linear Maxwell electrodynamics
whereas the same is not true for most of the NED models. In this work, since
we aim CF spacetimes, there will be no singularity problem to worry about.
We give exact CF metrics in all dimensions (with the exception of $n=4$) as
a power-law Lagrangian in the absence of a cosmological constant when the
source is electric type NED field. The case $n=4$ is added at the end of the
paper which is not of power-law form.

\paragraph{n+1-dimensional electric BR-solution in nonlinear Maxwell theory:}

The action of the theory is ($8\pi G=1=c$) 
\begin{equation}
I=\int \sqrt{-g}d^{n+1}x\left( R+\mathcal{L}\left( \mathcal{F}\right) \right)
\label{eq:1}
\end{equation}%
in which $R$ is the scalar curvature and $\mathcal{L}\left( \mathcal{F}%
\right) $ is the NED Lagrangian given by 
\begin{equation}
\mathcal{L}\left( \mathcal{F}\right) =\alpha \left( -\mathcal{F}\right) ^{s}.
\label{eq:2}
\end{equation}%
Herein, $\alpha $ is a coupling constant and $s$ is a real parameter. The
specific choice of $\left( -\mathcal{F}\right) ^{s}$ is due to our choice of
the pure electric field for the electromagnetic source, (see Eq. (\ref{eq:3}%
)) which causes $\mathcal{F}<0.$ Hence, $-\mathcal{F}>0$ does not impose any
additional restriction on $s$. The electromagnetic $2$-form $\mathbf{F}$ is
pure electric type given by 
\begin{equation}
\mathbf{F}=E\left( r\right) dt\wedge dr  \label{eq:3}
\end{equation}%
where $E\left( r\right) $ is the radial electric field depending only on
radial coordinate $r.$ Our CF line element is chosen to be 
\begin{equation}
ds^{2}=\phi ^{2}\left( r\right) \left( -dt^{2}+dr^{2}+r^{2}d\Omega
_{n-1}^{2}\right) .  \label{eq:4}
\end{equation}%
in which $\phi \left( r\right) $ depends also on $r$ only and together with $%
E\left( r\right) $ are to be determined. Furthermore, $d\Omega
_{n-1}^{2}=d\theta
_{1}^{2}+\sum\limits_{i=2}^{n-2}\prod\limits_{j=1}^{i-1}\sin ^{2}\theta
_{j}d\theta _{i}^{2}$ stands for the ($n-1$)-dimensional unit sphere line
element in which $0\leq $ $\theta _{k}$ $\leq \pi $ with $k=1..n-2$, and $%
0\leq \theta _{n-1}\leq 2\pi $. Being regular it may be used as a model to
represent inner geometry of an elementary particle \cite{7}. Herein, with
the exception of $n=4$ we show that higher dimensional such spacetimes are
constructed within the power-law NED theory. The interesting aspect of NED
is that the theory is still abelian so that its nonlinearity may account for
the similar aspects of the non-abelian Yang-Mills theory. We must admit,
however, that a disadvantage of the BR spacetime is that it lacks a center,
such as $r=0$, so that asymptotic freedom of particles are not satisfied.
These aspects can be seen from a detailed analysis of the time-like
geodesics. Being also not asymptotically flat, it fails to satisfy the
conservation laws that are derived in particular for such spacetimes. The
energy momentum tensor of NED is given by 
\begin{equation}
T_{\mu }^{\nu }=\frac{1}{2}\left( \mathcal{L}\delta _{\mu }^{\nu }-4\mathcal{%
L}_{\mathcal{F}}F_{\mu \lambda }F^{\nu \lambda }\right)  \label{eq:5}
\end{equation}%
in which $\mathcal{L}_{\mathcal{F}}$ stands for $\frac{\partial \mathcal{L}}{%
\partial \mathcal{F}}$. This energy-momentum tensor implies 
\begin{equation}
T_{t}^{t}=T_{r}^{r}=\frac{1}{2}\left( \mathcal{L}-2\mathcal{F}\mathcal{L}_{%
\mathcal{F}}\right)  \label{eq:6}
\end{equation}%
and 
\begin{equation}
T_{\theta _{i}}^{\theta _{i}}=\frac{1}{2}\mathcal{L}  \label{eq:7}
\end{equation}%
for $1\leq i\leq n-1.$ The corresponding NED equation takes the form 
\begin{equation}
d\left( \mathbf{\tilde{F}}\mathcal{L}_{\mathcal{F}}\right) =0  \label{eq:8}
\end{equation}%
in which $\mathbf{\tilde{F}}$ is the $(n-1)-$form dual of $\mathbf{F}$ given
by $\mathbf{\tilde{F}}=E\left( r\right) \left( r\phi \right)
^{n-3}r^{2}\left( \prod\limits_{k=1}^{n-2}\left( \sin \theta _{k}\right)
^{n-k-1}\right) d\theta _{1}\wedge d\theta _{2}\wedge d\theta _{3}\wedge
...d\theta _{n-1}$. The gravitational field equations read as 
\begin{equation}
G_{\mu }^{\nu }=T_{\mu }^{\nu }.  \label{eq:9}
\end{equation}%
The nonzero components of the Einstein's tensor in $n+1$-dimensions are
obtained to be%
\begin{equation}
G_{t}^{t}=\frac{n-1}{\phi ^{4}r}\left( r\phi \phi ^{\prime \prime }+\left(
n-1\right) \phi \phi ^{\prime }+\frac{n-4}{2}r\phi ^{\prime 2}\right) ,
\label{tt}
\end{equation}%
\begin{equation}
G_{r}^{r}=\frac{\left( n-1\right) \phi ^{\prime }}{\phi ^{4}r}\left( \left(
n-1\right) \phi +\frac{n}{2}r\phi ^{\prime }\right)  \label{rr}
\end{equation}%
and%
\begin{equation}
G_{\theta i}^{\theta _{i}}=\frac{n-1}{\phi ^{4}r}\left( r\phi \phi ^{\prime
\prime }+\left( n-2\right) \phi \phi ^{\prime }+\frac{n-4}{2}r\phi ^{\prime
2}\right) .  \label{thth}
\end{equation}%
With the constraint $T_{t}^{t}=T_{r}^{r}$ one imposes $G_{t}^{t}=G_{r}^{r}$
which amounts to%
\begin{equation}
\phi \phi ^{\prime \prime }=2\phi ^{\prime 2}.  \label{eq}
\end{equation}%
The latter yields an exact solution for $\phi \left( r\right) $ given by 
\begin{equation}
\phi \left( r\right) =\frac{\beta }{r+r_{0}}  \label{eq:10}
\end{equation}%
in which $\beta $ and $r_{0}$ are two integration constants. Note that in
analogy with the $3+1-$dimensional BR spacetime we set $r_{0}=0.$
Accordingly, the Einstein tensor components of the $n+1-$dimensional BR
spacetime (\ref{eq:4}) becomes 
\begin{equation}
G_{\mu }^{\nu }=-\frac{n-1}{2\beta ^{2}}diag\left( n-2,n-2,\underset{n-1%
\text{ times}}{\underbrace{n-4,n-4,...,n-4}}\right) .  \label{eq:11}
\end{equation}%
Now, the Einstein's field equations (\ref{eq:9}) reveal that 
\begin{equation}
\frac{\left( n-1\right) \left( n-2\right) }{2\beta ^{2}}=-\frac{1}{2}\left( 
\mathcal{L}-2\mathcal{FL_{F}}\right)  \label{eq:12}
\end{equation}%
and 
\begin{equation}
\frac{\left( n-1\right) \left( n-4\right) }{2\beta ^{2}}=-\frac{1}{2}%
\mathcal{L}.  \label{eq:13}
\end{equation}%
Knowing that $\mathcal{FL_{F}}=s\mathcal{L}$ one finds, from the combination
of (\ref{eq:12}) and (\ref{eq:13}) that the only consistent value for $s$ is
obtained to be 
\begin{equation}
s=-\frac{1}{n-4}.  \label{eq:14}
\end{equation}%
Furthermore, since\ $\mathcal{F}=2F_{tr}F^{tr}=-2\frac{E^{2}}{\phi ^{4}},$
the NED field equations i.e., (\ref{eq:8}) implies 
\begin{equation}
\frac{\alpha }{n-4}\left( 2\frac{E^{2}}{\phi ^{4}}\right) ^{-\frac{n-3}{n-4}%
}E\left( r\right) \left( r\phi \right) ^{n-3}r^{2}=C=\text{constant}
\label{ME}
\end{equation}%
where $C$ is an integration constant. Upon considering $\phi r=\beta $ the
latter equation implies%
\begin{equation}
E=\frac{q}{r^{2}}  \label{EF}
\end{equation}%
in which%
\begin{equation}
q=\left( \frac{n-4}{\alpha }C\right) ^{\frac{n-4}{2-n}}2^{\frac{n-3}{2-n}%
}\beta ^{\frac{n\left( n-3\right) }{n-2}}.  \label{CHARGE}
\end{equation}%
On the other hand, the angular component of the Einstein's field equation
i.e., Eq. (\ref{eq:13}), implies%
\begin{equation}
\frac{\left( n-1\right) \left( n-4\right) }{2\beta ^{2}}=-\frac{1}{2}\alpha
\left( 2\frac{q^{2}}{\beta ^{4}}\right) ^{-\frac{1}{n-4}}.  \label{Eq2}
\end{equation}%
Solving Eqs. (\ref{CHARGE}) and (\ref{Eq2}) simultaneously for $q$ and $%
\beta $ we obtain%
\begin{equation}
q=\frac{2^{\frac{n-3}{2}}C^{n-2}\left( \frac{\alpha }{4-n}\right) ^{\frac{%
\left( n-1\right) \left( n-4\right) }{2}}}{\left( n-1\right) ^{\frac{n\left(
n-3\right) }{2}}}  \label{eq:15}
\end{equation}%
and 
\begin{equation}
\beta =\sqrt{2}\left( \frac{\alpha }{4-n}\right) ^{\frac{n-4}{2}}\left(
n-1\right) ^{\frac{2-n}{2}}C.  \label{eq:16}
\end{equation}%
Let's add that the total physical charge is given by (for the details see 
\cite{CHARGE}) 
\begin{equation}
Q=\frac{1}{8\pi }\oint\nolimits_{S^{n-1}}\mathcal{L}_{\mathcal{F}}F^{\mu \nu
}dS_{\mu \nu }  \label{Charge1}
\end{equation}%
in which $Q$ is the electric charge and $dS_{\mu \nu }$ is the directed
surface element on the ($n-1$)-sphere $S^{n-1}$ with radius $\phi r=\beta .$
The explicit integration of (\ref{Charge1}) yields%
\begin{equation}
Q=\frac{\alpha }{4\pi \left( 4-n\right) }2^{\frac{n-3}{4-n}}\left( \frac{q}{%
\beta ^{2}}\right) ^{\frac{n-2}{4-n}}\beta ^{n-1}\omega _{n-1}
\label{Charge2}
\end{equation}%
where $\omega _{n-1}=2\pi ^{n/2}/\Gamma \left( n/2\right) $ is the volume of
the unit ($n-1$)-sphere and $\Gamma \left( n/2\right) $ stands for the gamma
function. In four dimensions where $n=3,$ $s=1$ and $\alpha =1$ one finds $%
Q=q=C=\beta $ as it is expected and the solution becomes the electric BR
spacetime in standard linear Maxwell theory where $\mathcal{L}=-\mathcal{F}$
. Let us note that, in (\ref{eq:15}) and (\ref{eq:16}), $\alpha $ is a
theory parameter whose sign should be fixed from the beginning, i.e., $%
\alpha >0$ for $n=2,3$ and $\alpha <0$ for $n>4.$ For $n=4$ the case is
considered separately at the end of the paper, since it is not of power-law
Lagrangian. In $2+1-$dimensions, $n=2,$ the Lagrangian becomes $\mathcal{L}%
=\alpha \sqrt{-\mathcal{F}}$ and so on.

It is worth also to find the energy-momentum tensor defined by $T_{\mu
}^{\nu }=diag\left( -\rho ,p_{r},p_{\perp },...,p_{\perp }\right) $ in which 
$\rho =$$\frac{\left( n-1\right) \left( n-2\right) }{2\beta ^{2}}\geq 0$ for 
$n\geq 2$, $p_{r}=-\rho $ and $p_{\perp }=-\frac{n-4}{n-2}\rho $ which imply
that $\rho +p_{i}\geq 0$. \ Hence, the weak energy conditions are satisfied,
indicating that the NED source provides a physical energy-momentum tensor.
With reference to the energy conditions in arbitrary dimensions for Type-I
energy momentum tensors in Ref. \cite{Maedaa} we can easily show that energy
conditions are satisfied (see also \cite{8} for $4$-dimensions). The strong
energy condition, for instance, which amount to $\rho +p_{i}\geq 0$ and $%
\left( n-3\right) \rho +\left( n-1\right) p_{\perp }\geq 0$ are satisfied.

Furthermore, to see the topology of the higher dimensional BR we employ the
transformation%
\begin{equation}
t=\frac{\beta \sqrt{\beta ^{2}+R^{2}}\sin \left( \frac{T}{\beta }\right) }{R+%
\sqrt{\beta ^{2}+R^{2}}\cos \left( \frac{T}{\beta }\right) }  \label{Tran1}
\end{equation}%
and%
\begin{equation}
r=\frac{\beta ^{2}}{R+\sqrt{\beta ^{2}+R^{2}}\cos \left( \frac{T}{\beta }%
\right) }  \label{Trans2}
\end{equation}%
so that the higher dimensional BR solution (\ref{eq:4}) is transformed as%
\begin{equation}
ds^{2}=-\left( 1+\frac{R^{2}}{\beta ^{2}}\right) dT^{2}+\left( 1+\frac{R^{2}%
}{\beta ^{2}}\right) ^{-1}dR^{2}+\beta ^{2}d\Omega _{n-1}^{2}  \label{Trans3}
\end{equation}%
which shows that its topology is the direct product of $AdS_{2}\times
S^{n-1}.$ In other words, geometrically, the higher-dimensional BR solution
is the direct product of two constant curvature manifolds, a $2$-dimensional
Anti-de-Sitter\ and an ($n-1$)-sphere of constant radius $\beta .$ Let's add
that, the higher dimensional BR spacetime is important because it represents
the near horizon geometry of the higher-dimensional Reissner-Nordstr\"{o}m
black hole (see, for example, Ref. \cite{Lemos,NH}). Furthermore, the $3+1$%
-dimensional BR solution is known to be the unique CF Einstein-Maxwell
solution. It remains to be seen whether the similar uniqueness is valid also
for NED and in higher dimensions.

\paragraph{The electric BR solution in $5$-dimensions}

Our higher-dimensional BR solution given in previous section fails for $%
n+1=5 $ dimensions. Here in this section we give the alternative Lagrangian
which admits the similar solution in five dimensional spacetime. The NED
Lagrangian (\ref{eq:2}) has to be modified as%
\begin{equation}
\mathcal{L}\left( \mathcal{F}\right) =\alpha \ln \left( \frac{\mathcal{F}}{%
\mathcal{F}_{0}}\right)
\end{equation}%
in which $\alpha $ and $\mathcal{F}_{0}$ are two dimensionful constants.
Knowing that a constant term in $\mathcal{L}\left( \mathcal{F}\right) $ can
be interpreted as a cosmological constant we set $\mathcal{F}_{0}$ to have
physically the same unit as $\mathcal{F}$ with a magnitude of $-1$. Hence,
we continue with the choice 
\begin{equation}
\mathcal{L}\left( \mathcal{F}\right) =\alpha \ln \left( -\mathcal{F}\right) .
\label{LAG}
\end{equation}%
The CF line element is given by%
\begin{equation}
ds^{2}=\phi ^{2}\left( r\right) \left( -dt^{2}+dr^{2}+r^{2}\left( d\theta
_{1}^{2}+\sin ^{2}\theta _{1}\left( d\theta _{2}^{2}+\sin ^{2}\theta
_{2}d\theta _{3}^{2}\right) \right) \right)
\end{equation}%
which yields the dual field of the electric field (\ref{eq:3}) as%
\begin{equation}
\mathbf{\tilde{F}}=E\left( r\right) \phi \left( r\right) r^{3}\sin
^{2}\theta _{1}\sin \theta _{2}d\theta _{1}\wedge d\theta _{2}\wedge d\theta
_{3}.
\end{equation}%
The Einstein's field equations are given by Eq. (\ref{eq:9}) which due to
the fact that $T_{t}^{t}=T_{r}^{r}$ we impose again $G_{t}^{t}=G_{r}^{r}$
which amounts to Eq. (\ref{eq:10}). Again to get the usual form of the BR
solution we set $r_{0}=0.$ On the other hand, the Maxwell NED field equation
(\ref{eq:8}) becomes%
\begin{equation}
d\left( \frac{\alpha }{\mathcal{F}}E\left( r\right) \phi \left( r\right)
r^{3}\sin ^{2}\theta _{1}\sin \theta _{2}d\theta _{1}\wedge d\theta
_{2}\wedge d\theta _{3}\right) =0.
\end{equation}%
and upon knowing $\mathcal{F}=2F_{tr}F^{tr}=-\frac{2E^{2}}{\phi ^{4}}$ which
is only a function of $r,$ it amounts to%
\begin{equation}
\frac{\phi ^{5}}{E}r^{3}=C=\text{constant}
\end{equation}%
where $C$ is an integration constant. Here upon imposing $\phi =\frac{\beta 
}{r}$ one finds $E=\frac{q}{r^{2}}$ in which $q=\frac{\beta ^{5}}{C}$.
Finally considering the explicit form of the Einstein's tensor given by (\ref%
{eq:11}), i.e., 
\begin{equation}
G_{\mu }^{\nu }=-\frac{3}{\beta ^{2}}diag\left( 1,1,0,0,0\right)
\end{equation}%
and the energy-momentum tensor (\ref{eq:6}) and (\ref{eq:7}) one finds%
\begin{equation}
T_{\theta _{i}}^{\theta _{i}}=\frac{1}{2}\alpha \ln \left( -\mathcal{F}%
\right) =0
\end{equation}%
which results in $\frac{2E^{2}}{\phi ^{4}}=1.$ This in turn gives 
\begin{equation}
q^{2}=\frac{\beta ^{4}}{2}.
\end{equation}%
On the other hand the $rr$ or $tt$ component of the Einstein equation yields%
\begin{equation}
\beta ^{2}=-\frac{3}{\alpha }  \label{alpha}
\end{equation}%
and consequently $q^{2}=\frac{9}{2\alpha ^{2}}.$ Eq. (\ref{alpha}) implies
that $\alpha $ as a theory parameter should be negative. Finally, integral (%
\ref{Charge1}) reveals the electric charge given by%
\begin{equation}
Q=\frac{\pi \alpha \beta ^{3}}{4q}.
\end{equation}%
This completes the solution for the $5$-dimensional NED coupled BR solution
that was missing in the power-law Lagrangian.

Finally we discuss the causality and stability of our model of NED. The
causality imposses the speed of sound $c_{s}$ to be less than the speed of
light, i.e., $c_{s}\leq 1.$ For classical stability analysis we refer to 
\cite{CAS1,CAS2,CAS3} in which the perturbed background energy density is
given by $\rho _{B}\left( t,\mathbf{r}\right) =\rho _{B}\left( t\right)
+\delta \rho _{B}\left( t,\mathbf{r}\right) $. The wave equation for small
perturbations satisfies%
\begin{equation}
\delta \ddot{\rho}_{B}=c_{s}^{2}\triangledown ^{2}\delta \rho _{B}
\end{equation}%
where the dot implies time derivative and $c_{s}^{2}\geq 0$. The square of
the speed of sound in our case is given by%
\begin{equation}
c_{s}^{2}=\frac{\partial p_{av}}{\partial \rho }=-\frac{n^{2}-4n+2}{n\left(
n-2\right) }
\end{equation}%
in which the average pressure is%
\begin{equation}
p_{av}=\frac{p_{r}+\left( n-1\right) p_{\perp }}{n}.
\end{equation}%
It is seen that for $n=3$ we obtain $c_{s}^{2}=\frac{1}{3},$ which satisfies
both causality and stability conditions. For $n\neq 3,$ however, these
conditions are violated.

In summary, we started initially with an arbitrary power-law form of NED
Lagrangian and we ended up with the specific power related to the
dimensionality of the conformally flat spacetime in arbitrary dimensions. It
is clear from (\ref{eq:14}) that for $n=4$, i.e., in $4+1$-dimensions, the
power-law Lagrangian does not admit a CF solution. The alternative form of
the Lagrangian which yields a CF spacetime in this specific five-dimensions
is obtained to be of the Logarithmic form given in Eq. (\ref{LAG}). Our
higher dimensional CF solution coincides with the well-known BR solution for 
$n+1=4$ which suggests that we call the general solution to be the
higher-dimensional BR spacetime. We have shown that the energy-conditions
are satisfied in analogy with the $4$-dimensional BR solution.

\end{document}